\documentclass[prd,twocolumn,superscriptaddress,altaffilletter,amssymb,showpacs,nofootinbib]{revtex4}
\usepackage[dvips]{graphicx}
\usepackage{amsmath}
\usepackage[latin1]{inputenc}
\newcommand{\be}{\begin{equation}}
\newcommand{\ee}{\end{equation}}
\newcommand{\bea}{\begin{eqnarray}}
\newcommand{\eea}{\end{eqnarray}}

\PassOptionsToPackage{linktocpage}{hyperref}
\usepackage[draft=false]{hyperref}
\usepackage{threeparttable}
\usepackage[draft=false]{hyperref}
\usepackage{threeparttable}

\begin{document}

\title{Born-Infeld type modification of  the gravity}

\author{Dagoberto Escobar}\email{dagoberto.escobar@reduc.edu.cu} 
\affiliation{Departamento de F\'isica, Universidad de Camagüey, A.P. 74650, Cuba.}

\date{\today}

\begin{abstract}
We study a non-linear modification to General Relativity in
which the standard Einstein-Hilbert action is replaced by a 
Born-Infeld type action. Also study us stability issues
to judge about viability of this modification.
We establish the conditions that this modification must satisfy for to avoid the problems associated with the Dolgov-Kawasaki 
instability, tachyon instability and negative effective gravitational coupling.
 The particle content of 
gravitational spectrum of the linearized Born-Infeld
theory shows the existence of massless gravitons plus new degree
of freedom of $0$-spin associated with $R^2$ term. For a toy model we proved that for appropriate values 
of the background curvature, this theory  is free of  ghost, Dolgov-Kawasaki instability and tachyon instability also it has a positive effective gravitational coupling. We find the scalar-tensor theory  equivalent at this $f(R)$-theory in the Einstein frame, and we study some properties of the scalar potential.  
\\

\end{abstract}

\maketitle

\section{Introduction}
Recently the cosmological observations Supernova (SNIa) \cite{Riess:1999ti}, large scale structure \cite{Tegmark2006}, cosmic microwave background \cite{Peiris:2003ff}, the integrated Sachs-Wolfe effect \cite{Vielva2006}, baryonic acoustic oscillations \cite{Eisenstein:2005su} and from gravitational lensing \cite{Contaldi2003}  have offered strong evidence
that the expansion of our universe is accelerating. The General Relativity (GR) cannot explain the acceleration of late universe. In most investigations the origin of the acceleration is attributed to a mysterious component with a negative pressure, called Dark Energy (DE). The preferred candidate for
this entity is a cosmological constant $\Lambda$. The Cosmological Constant problem has been a longstanding problem in fundamental theoretical physics \cite{Weinberg:1988cp}. An alternative at Dark Energy is modified the Einstein gravity theory. Several theories of gravity of this kind have been proposed since long ago in \cite{Eddington1924}.
The modifications of gravity have been used in many contexts, perhaps the most famous is the first  model of inflation introduced by Starobinsky \cite{Starobinsky:1980te}.
However, in the last few years, modifications of gravity have been attracting attention as an alternative to the Dark Energy \cite{Capozziello2002,Capozziello2003g,Carroll:2003wy}, in order to explain the late-time acceleration at large cosmological scales \cite{Amendola2007c,Carroll:2006jn,Nojiri2005b}.
 
The General Relativity is based in the Einstein-Hilbert action (EH)
\be\label{EH}
S_{EH}=\frac{1}{2\kappa^{2}}\int d^4x\sqrt{-g}\left(R-2\Lambda\right)
\ee
Where $\kappa^{2}=8\pi G$ is the gravitational coupling constant, $R$ is the curvature scalar, $g$ is the determinant of the metric and $\Lambda$ the cosmological constant.
Numerous modification to the classical EH action have been proposed motivated for different reasons. In particular, renormalization at one-loop
demands that the EH action be supplemented by higher order curvature
terms \cite{Utiyama1962}\footnote{Higher order actions are indeed renormalizable but not unitary \cite{Stelle:1976gc}.}. Besides, when quantum corrections or string theory are taken into account, the
effective low energy action for pure gravity admits higher order curvature invariants \cite{Birrell1982}.
The most well-known alternative to general relativity, are Scalar-Tensor Theories \cite{Brans:1961sx,Bergmann:1968ve}, but there are still numerous proposals for modified gravity in contemporary literature. Typical examples are the braneworld gravity of  Randall-Sundrum \cite{Randall:1999vf,Randall:1999ee} or Dvali-Gabadadze-Porrati \cite{Dvali:2000hr}, quantum gravity \cite{Reuter:2012id}.
But there is other way to modify the  Einstein gravity theory. 
If the curvature is replaced in \eqref{EH} by an arbitrary function $f(R)$, the  gravity action is given by 
\be\label{fr}
S_g=\frac{1}{2\kappa^{2}}\int d^4x\sqrt{-g}f(R)
\ee
This kind of modification to GR has been extensively in the literature.
We can consider a series expansion of $f(R)$ in the following form
\be\label{power}
f(R)=....+\frac{\mu_2}{R^2}+\frac{\mu_1}{R}+R+\frac{R^2}{\eta_2}+\frac{R^3}{\eta_3}+.....
\ee
where  $\mu_i$ and $\eta_i$ are constants with the appropriate dimensions.
The expansion \eqref{power}  includes  fenomelogically interesting term \cite{Sotiriou2008b}. The positive power  in \eqref{power} modified gravity at high energy (early universe), for example,  the model $f(R)=R+\beta R^2 (\beta>0)$  can lead to the inflation in the early universe because of the presence of the $\beta R^2$ term. In fact, this is the model of inflation proposed
by Starobinsky \cite{Starobinsky:1980te}. The negative power in \eqref{power} modified gravity at lower energy (late universe), the model $f(R)=R+\frac{\alpha}{R^n} $ with $(\alpha>0,n>0)$ is a simplest modified gravity scenario to realize the late-time acceleration. However it was proved that
this model is plagued by a matter instability \cite{Faraoni2006c,Dolgov2003}, as well as by a difficulty to satisfy local
gravity constraints \cite{Olmo2005,Olmo2005c,Faraoni2006b}.
The interesting feature of the theories
with negative powers of curvature is that they may be expected from
some time-dependent compactification of string/M-theory as it was demonstrated
in \cite{Nojiri:2003rz}.
The action with negative and positive power of the curvatures has been studied in \cite{Nojiri:2003ft}.

The $f(R)$ theories have been studied in many contexts inflation \cite{Starobinsky:1980te}, Dark Energy \cite{Nojiri:2006su,Nojiri:2007as,Nojiri2008a,Nojiri:2003ni}, large scale structure \cite{Song:2006ej}.
Even if $f(R)$ theories were not a
viable alternative to explain late-time acceleration of the
expansion, these modification are of great relevance to study early-time inflation
\cite{Starobinsky:1980te}.

We can add matter fields to $f(R)$ gravity action \eqref{fr}, in this case the total action  is given by   
\be\label{frM} 
S=\frac{1}{2\kappa^{2}}\int d^4x\sqrt{-g}f(R)+S_m(g_{\mu\nu},\xi)
\ee
where $\xi$ represent the matter fields, i.e. electromagnetic field. Varying  the action \eqref{frM}  with respect to the metric, we find the following field equation \footnote{The field equation in $f(R)$ gravity also can be obtained through the known Palatini formalism \cite{Sotiriou2008b}. In Palatini formalism the motion equations are of second order (like usual Einstein equation), while that the motion equations \eqref{FE} are of fourth order with respect to metric.}  
\be\label{FE}
f'(R) R_{\mu\nu}-\nabla_{\mu}\nabla_{\nu}f'(R)+g_{\mu\nu}\Box f'(R)
-\frac{1}{2}g_{\mu\nu}f(R)=\kappa^2T_{\mu\nu}^{(m)}
\ee
where 
\be\label{matt}
T_{\mu\nu} ^{(m)}=-\frac{2}{\sqrt{-g}}\frac{\delta S_m}{\delta g_{\mu\nu}}
\ee
is the momentum-energy tensor of the matter.
The field equation \eqref{FE} can be written in the following form
\be G_{\mu\nu}=\kappa^2 _{eff}\left(T_{\mu\nu}^{(m)} +T_{\mu\nu}^{(cur)}\right)
\ee
where $G_{\mu\nu}$ is the Einstein tensor, $T_{\mu\nu}^{(cur)}$ is the momentum energy tensor of curvature
\bea
T_{\mu\nu}^{cur}=\frac{1}{\kappa^2}\left(\frac{f(R)-Rf'(R)}{2}g_{\mu\nu}+\nabla_{\mu}\nabla_{\nu}f'(R)\right)\\
-\frac{g_{\mu\nu}}{\kappa^2}\Box f'(R)\nonumber
\eea
and effective gravitational coupling
\be\label{coup}
\kappa^2 _{eff}=\frac{\kappa^2}{f'(R)}
\ee
The trace of the field equation \eqref{FE}
\be\label{trace}
f'(R)R-2f(R)+3\Box f'(R)=\kappa^2 T
\ee
The equation \eqref{trace} relates $R$ with $T$ differentially and not algebraically as occur in the GR. This means that the field equation in $f(R)$-theories  admit a larger variety of solutions than GR.

The $f(R)$ theories are excellent candidates to gravity theory. The $f(R)$ theories include some of the basic characteristic of higher-order gravity theories. Also there are several reasons to believe that $f(R)$ theories are  only of the higher-order gravity theories, that it can avoid the well-known Orstrogradski instability \cite{Woodard:2006nt}.
\section{Born-Infeld type modification}
There are several ways to modify the EH action.
In this paper, we consider a modification  based on the gravitational analogues of non-linear Born-Infeld electrodynamics \cite{Born1934},  modification for smoothing out singularities \cite{Carroll:2004de,Carroll:2003wy,Capozziello2003d}\footnote{ In \cite{Nojiri:2003ft} has been studied how to avoid the initial as well as final singularities in
modified gravity.}. 
Through this modification the Einstein-Hilbert Lagrangian
density ${\cal L}_{EH}=\sqrt{-g}L$ can be replaced by one of the Born-Infeld form:
\be\label{LBI}
{\cal L}_{EH}\rightarrow{\cal L}_{BI}=\sqrt{-g}\lambda\left(1-\sqrt{1+\frac{2L}{\lambda}}\right)
\ee
where the constant $\lambda$ is associated with the curvatures scale accessible
to the theory.
Also we can replace the Lagrangian $L$ in \eqref{LBI} by a $f(R)$ function within the square root (see \cite{Quiros:2010vz})
\be\label{LBIFR}
{\cal L}_{BI}=\sqrt{-g}\lambda\left(1-\sqrt{1+\frac{2f(R)}{\lambda}}\right)
\ee
Let us consider a function $f(R)\propto R^n$ in BI-type Lagrangian density
\eqref{LBIFR} for gravity, in this case the action of the pure gravity \eqref{fr} can be rewritten as 
\be\label{BI}
S_{BI}=\frac{1}{2\kappa^{2}}\int d^4x\sqrt{-g}\left(1-\sqrt{1+\epsilon\alpha R^n}\right) 
\ee
where $\epsilon=\pm1$, the constant $\alpha$ is a constant with mass dimension  and $(n>0)$.

Other variants of Born-Infeld
type gravity was considered in \cite{Deser1998,Comelli2005a,Banados:2008fj,Pani:2012qb}. The BI gravity has been intensively studied in many context, as cosmological framework proves to lead to interesting behavior \cite{Fiorini:2009ux,Garcia-Salcedo2000,Salcedo2010a}.  In particular has been analyzed the black hole properties \cite{ Chemissany:2008fy,Bostani:2009tb,Dehghani:2008qr,Ghodsi:2010ev,Miskovic:2008ck}, the astrophysical phenomenology
\cite{Avelino:2012ge,Lubo:2008db}, dark energy phenomenology \cite{Banados:2008rm}, massive gravity \cite{Gullu:2010pc} and primordial inflation \cite{Tolley:2008na}.

Born-Infeld gravity can be phenomenologically viable, if this
modification satisfy several physically 
motivated requirements.
\begin{itemize} 
\item Reduction to Einstein-Hilbert action  at small curvature.
\item Ghost-free
\item Regularization of singularities, i.e. Schwarzschild
singularity
\item Supersymmetrizability\footnote{This requirement results quite stringent and is probably
implemented if gravity descends from String Theory \cite{Wohlfarth2004,Gates2001}.}
\end{itemize} 
The function $f(R)$ that appear in the action \eqref{BI} also satisfies the condition $\lim_{R\rightarrow0} f(R)\rightarrow0$ 
which means that there is a flat space-time solution.
 
\subsection{Born-Infeld Lagrangian expansion}
In this section we study
the expansion around maximally symmetric vacuum
spaces of constant curvature of the $f(R)$-theory that appear in \eqref{BI}. In the next section we will analyze the stability issues in these constant curvature spaces. 

The equation \eqref{trace} for $R=Const$ and $T=0$ reduce to 
\be\label{trace1}
f'(R)R-2f(R)=0
\ee
which is an algebraic equation in $R$.

For the action \eqref{BI} the roots of the equation \eqref{trace1} are
\be\label{root}
R_0=0 \qquad R_0=\left[\frac{8(n-2)}{\epsilon\alpha(n-4)^2}\right]^{\frac{1}{n}}
\ee
For $R=0$ the equation \eqref{FE} reduced to $R_{\mu\nu}=0$  and the maximally symmetric solution is Minkowski spacetime. The  second root \eqref{root} reduced to $R_{\mu\nu}=\frac{R_0}{4}g_{\mu\nu}$ the equation \eqref{FE} and the maximally symmetric solution is de Sitter space if $R_0>0$ or Anti de Sitter ($AdS$) space if $R_0<0$.

In an empty space of constant curvature $R_0$ all $f(R)$ theories admit Schwarzschild-de Sitter(or $AdS$) solution \cite{Nojiri:2010wj}. We can find static solutions with spherical symmetric of the theory \eqref{BI} in the Schwarzschild form
\be ds^2=-A(r)dt^2+A^{-1}(r)dr^2+r^2\left(d\theta^2+\sin^2\theta d\varphi^2\right)
\ee
where
\be
A(r)=1-\frac{2MG}{r}-\frac{1}{12}\left[\frac{8(n-2)}{\epsilon\alpha(n-4)^2}\right]^{\frac{1}{n}} r^2
\ee 
In above solution we can see that AdS-Schwarzschild black holes in Einstein gravity are also solutions of  Born-Infeld gravity \eqref{BI}.

Let us consider the expansion of the  action \eqref{BI} around
maximally symmetric vacuum spaces of constant curvature
$R = R_0$, in the neighborhood of the point $R = R_0$.
\be\label{taylor}
f(R)=f(R_0)+\sum^{\infty}_{n=1}\frac{f^{(n)}(R_0)}{n!}(R-R_0)^n
\ee
If we consider until second order term $(R-R_0)^2$ in the series \eqref{taylor}, one has 
\bea 
f(R)=f(R_0)+f'(R_0)(R-R_0)\\
+\frac{1}{2}f''(R_0)(R-R_0)^2+\mathcal{O}(R^3)\nonumber
\eea
the above expansion can be written as
\be\label{serie}
f(R)=\beta\left(-2\Lambda+R+\frac{R^2}{6m_0 ^2}\right)
\ee 
where we have considered the following notation 
\bea\label{mass}
-2\Lambda=\frac{f_0-R_0 f_0'+\frac{R_0 ^2}{2}f_0'}{f'-R_0f_0''}\qquad\beta=f'-R_0f_0''\\
 m_0 ^2=\frac{f'-R_0f_0''}{3f_0''}\nonumber
\eea
and
\be
f_0=f(R_0)\quad  f_0'=\frac{d f}{d R}|_{R_0}. 
\ee
From the definition of the cosmological constant \eqref{mass} we find the expected
value $\Lambda=\frac{R_0}{4}$.
The linearized action is given by 
\be\label{BIlinea}
S=\frac{\beta}{2\kappa^2}\int d^4x\sqrt{-g}\left(-2\Lambda+R+\frac{R^2}{6m_0 ^2}\right)
\ee
Where the linearized theory is defined by the $\tilde{f}(R)$-theory
\be\label{fRlinea}
\tilde{f}(R)=R+\frac{R^2}{6m_0 ^2}-2\Lambda
\ee
For small curvatures the action \eqref{BIlinea} reduce to EH action \eqref{EH}. 
The model \eqref{fRlinea} was motivated to be renormalizable, results in the self-consistent inflation. Also the $R^2$ term in \eqref{BIlinea}, prevents from the singular behavior in the past and in the future \cite{Appleby:2009uf}.
The gravitational spectrum of the theory \eqref{BIlinea} shows the existence of massless gravitons relate with the $R$-term in the linearized action \eqref{BIlinea} plus new degree of freedom of $0$-spin and mass $m_0$ associated with the $R^2$-term in \eqref{BIlinea}. The mass of the scalar degree of freedom $m_0 ^2$ is  obtained in the weak-field limit, if we consider   small, spherically symmetric, perturbation on a  Sitter space of constant curvature \footnote{This new scalar degree of freedom also is called scalaron.}.
 
The exchange of this scalar field between two test particles change the Newtonian potential (see \cite{Faraoni1999f}) 
\be\label{new.pot}
\frac{1}{r}\rightarrow\frac{1}{r}\left(1+\frac{1}{3}e^{-m_0 r}\right)
\ee
For large values of $m_0$ the exponential term vanish in \eqref{new.pot} and we recover the Newtonian potential, however for small values  of $m_0$ the exponential term do not vanish and the corrections to Newton's law are not negligible.
\subsection{Stability criteria}\label{stability}
The study of the stability in higher order modification of GR is very important to judge about
viability of these modifications.
These theories are
plagued by several kinds of instabilities,  which lead to destabilize the theory\footnote{In \cite{Quiros:2010vz} was studied the stability requirement for Born-Infeld gravity with including Gauss-Bonnet term.}. For example, we have the Ostrogradski instability, based on the Ostrogradski's 
theorem \cite{Woodard:2006nt}.
This instability is associate with Lagragian that depend higher order time derivate. In this case that dependence cannot be eliminated by partial integration.
In \cite{Woodard:2006nt} proved that $f(R)$ theories are the only Ostrogradski-stable higher order modification to GR.
 
Other  instability is  known as Dolgov-Kawasaki
instability \cite{Dolgov2003}\footnote{This instability is also known as matter instability, in this paper we use Dolgov-Kawasaki instability.}.
This instability was discovered by Dolgov and Kawasaki in the theory $f(R)=R-\frac{\mu^4}{R}$

Dolgov and Kawasaki discovered this instability in the theory $f(R)=R-\frac{\mu^4}{R}$, this instability manifests itself on an extremely short time scale. The existence of this instability is sufficient to rule out a model \cite{Dolgov2003}. In \cite{Nojiri:2003ft,Nojiri:2003ni} proved that adding an $R^2$ term at this theory, it can removes this instability. The study of Dolgov-Kawasaki instability was generalized for arbitrary $f(R)$ theories in \cite{Faraoni2006c}. 

Other instabilities are caused by the presence
of a $0$-spin tachyon degree of freedom. This instability is associated with negative
values of the mass squared $m^2 _0$ defined in \eqref{mass}.
This new  scalar degree of freedom $(\phi=f'(R))$ with mass $m_0$ is associated to $R^2$ term in \eqref{BIlinea}.

From \eqref{serie}, we can see that the constant $\beta=f'-R_0f_0''$ is
an  factor that multiplies the action \eqref{BIlinea},
and, hence, it can change the sign of this, therefore is necessary that $\beta>0$.

Therefore, a $f(R)$ theory physically viable should be free of ghost, tachyon, Dolgov-Kawasaki instability, and negative effective gravitational coupling. 
 \begin{itemize}
\item Ostrogradski Stability

How we said previously the $f(R)$-theories are Ostrogradski-stable \cite{Woodard:2006nt}. 
Obviously the modification \eqref{BI} is Ostrogradski-stable.

\item Dolgov-Kawasaki Stability 
  
Assuming a positive effective gravitational coupling \eqref{coup}, we find
\be
\frac{dG_{eff}}{dR}=-\frac{\kappa^2f''(R)}{8\pi f'(R)^2}
\ee
when $f''< 0$, the effective gravitational coupling
 increases with the curvature. 
If $f''<0$ this instability can be seen as an instability in the gravity sector. In other words, this positive mechanism leads to destabilize
the theory \cite{Sotiriou2008b}. The modification \eqref{BI}
  avoid the Dolgov-Kawasaki instability if $f_0''\geq0$. 
\be\label{Riccicond} 
f_0''=-\frac{\alpha \epsilon  n R_0^{n-2} \left[\alpha \epsilon  (n-2) R_0^n+2 (n-1)\right]}{4 \left(\alpha \epsilon  R_0^n+1\right){}^{3/2}}\geq0
\ee
This condition also guarantees the classical stability of Schwarzschild black hole \cite{Seifert:2007fr}.

\item Positive Effective Gravitational Coupling

The effective gravitational coupling multiplies the linearized BI action \eqref{BIlinea}, is necessary that this factor do not change the sign of the action.
\be\label{EGC}
\frac{\beta}{2\kappa^2}=\frac{\alpha \epsilon  n R_0^{n-1} \left[\alpha \epsilon  (n-4) R_0^n+2 (n-2)\right]}{8\kappa^2\left(\alpha \epsilon  R_0^n+1\right){}^{3/2}}>0
\ee

\item Absence of Tachyon Instability

The Dolgov-Kawasaki stability condition ($f_0''>0$) and a positive
effective gravitational coupling ($f'-R_0f_0''>0$) together imply absence of tachyon in Born-Infeld  gravity.
The mass of the scalar degree of freedom for the theory \eqref{BI},
in the linear approximation is given by
\be
m_0 ^2=\frac{R_0\left[4-2n-\alpha \epsilon  (n-4) R_0^n\right]}{3 \alpha \epsilon  (n-2) R_0^n+6 (n-1)}>0
\ee

\item Ghost free 

Ghosts are massive state of negative norm that cause apparent absence of unitarity, it appear constantly in higher order gravity theories. A viable gravity theory should be ghost-free.  The stability condition $f''(R)\geq0$ essentially amounts guarantee that the scalar degree of freedom is not a ghost  \cite{Ferraris:1988zz,Stelle:1977ry,Utiyama1962}. If the BI modification is free of Dolgov-Kawasaki instability also is ghost-free.
\end{itemize}

\subsection{Scalar-Tensor theory equivalent to Born-Infeld gravity }
The $f(R)$ theory in the metric formalism can be cast in the form of Brans-Dicke theory \cite{Brans:1961sx} with a potential for the effective scalar degree of freedom \cite{Sotiriou2008b}. By introducing the auxiliary field $\chi$, we rewrite the action \eqref{BI} of the $f(R)$-gravity in the following form:
\be\label{AFR}
S_{BI}=\frac{1}{2\kappa^{2}}\int d^4x\sqrt{-g}\left(f'(\chi)(R-\chi)+f(\chi)\right) 
\ee
Varying this action with respect to $\chi$, we obtain
\be
f''(\chi)(R-\chi)=0
\ee
Provided $f''(\chi)\neq0$ it follows that $\chi=R$. Hence the action \eqref{AFR} recovers the action \eqref{BI} in $f(R)$ gravity. 
Using the conformal transformation $\hat{g}_{\mu\nu}=e^{-\phi}g_{\mu\nu}$ in the action \eqref{BI} where the new scalar degree of freedom is defined by
\be\label{field}
\phi=-\ln f'(R)
\ee
Usually in \eqref{field} is assumed  $f'(R)>0$, even if $f'(R)<0$ can be defined $\phi=-\ln\left|f'(R)\right|$. Then the sign in front of the scalar curvature becomes negative. In other words, the regimen of anti-gravity can be realized, however for the action \eqref{BI} the regimen anti-gravity $f'(R)<0$ can be avoid if $\epsilon=-1$.

The Einstein frame action looks like
\be\label{BI5}
S_{BI}^{E}=\int d^4x\sqrt{-\hat{g}}\left(\frac{1}{2\kappa^2}\hat{R}-\frac{3}{2}\left(\hat{\nabla}\phi\right)^2-V(\phi)\right) 
\ee
where $\left(\hat{\nabla}\phi\right)^2=\hat{g}^{\mu\nu}\partial_\mu\phi\partial_\nu\phi$ and potential is given by
\be
V(\phi)=\frac{f'(R)R-f(R)}{2\kappa^2f'(R)^2}| _{R=R(\phi)}
\ee
For the $f(R)$-function that appear in \eqref{BI} we find
\be\label{1.2}
e^{-\phi}=-\frac{n\epsilon\alpha R^{n-1}}{\sqrt{1+\epsilon\alpha R^n}}
\ee
If $R$ is large and using $\phi=-\ln\left|f'(R)\right|$ for $\epsilon=1$ we have
\be
e^{-\phi}\cong n\sqrt{\alpha}R^{\frac{n-2}{2}}
\ee
Then if $e^{-\phi}\rightarrow\infty$
\be\label{potential3}
V(\phi)=(\beta-\sqrt{\alpha\beta^n})e^{\frac{(n-4)}{n-2}\phi}
\ee
where $\beta=(\frac{1}{n\sqrt{\alpha}})^{\frac{2}{n-2}}$. The above potential has been found under approximation $1\ll\alpha R^n$.
For $0<n<2$ and $n>4$ the potential \eqref{potential3} diverge if $\phi\rightarrow\infty$ and $V(\phi)\rightarrow0$ if $\phi\rightarrow-\infty$.

How we can see in \eqref{1.2} to obtain the scalar-tensor theory equivalent to \eqref{BI} is still very complicated.
That represent a difficulty to study the particle spectrum of the theory \cite{Chiba:2005nz}.
However we can begin with the  action \eqref{BIlinea}, in this case the scalar degree of freedom is given
$\phi=-\ln\tilde{f}'(R)$
\be
e^{-\phi}=1+\frac{R}{3m_0 ^2}
\ee
with the following scalar  potential 
\be\label{potential1}
V(\phi)=\frac{3m_0 ^2}{2}\left(1-e^{-\phi}\right)^2+2\Lambda e^{-2\phi}
\ee
The above potential correspond to linearization of the original theory
and not the theory itself. However,  we
can prove, that the scalar-tensor theory dual to linearized theory \eqref{BIlinea} and the corresponding to original theory \eqref{BI} are equivalent.
This affirmation rest on the compute of the effective
mass of the scalar degree of freedom corresponding to theory \eqref{BIlinea}
\be
m_{eff}^2=\frac{\tilde{f}_0'-R_0\tilde{f}_0''}{3\tilde{f}_0''}
\ee
For the function $\tilde{f}(R)$ defined in \eqref{fRlinea}
\be
\tilde{f}_0'=1+\frac{R_0}{3m_0 ^2}\quad \tilde{f}_0''=\frac{1}{3m_0 ^2}
\ee
then the effective mass of the scalar degree of freedom
coincides with the mass of the $0$-spin excitation $m_0 ^2$
computed through \eqref{mass}: $m_{eff}^2=m_{0}^2.$

In the  Fig.1 we illustrate the potential \eqref{potential1}, in the this figure we can see that this potential 
when $\phi\rightarrow\infty$  $V(\phi)\rightarrow\frac{3m_0 ^2}{2}$, and when $\phi\rightarrow-\infty$ the potential diverge. 
\begin{figure}
\begin{center}
\includegraphics[height=2in,width=2in]{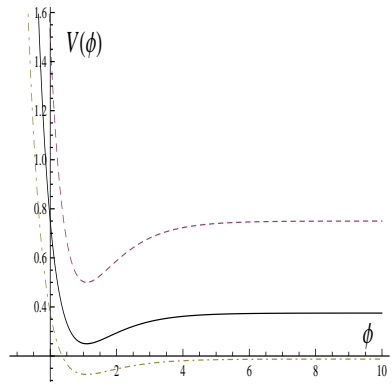}
\caption{Potential of the scalar field \eqref{potential1} corresponding to linearized theory for different choice of the parameters $m_0 ^2$ and $\Lambda$,  black line  $m_0 ^2=1/4, \Lambda=3/8$, dashed line $m_0 ^2=1/2, \Lambda=3/4$ and dotdashed line $m_0 ^2=1/8, \Lambda=3/16$.}
\end{center}
\end{figure}\label{fig1}
In the regimen $\phi\rightarrow\infty$ the potential is nearly constant $V(\phi)\cong\frac{3m_0 ^2}{2}$ which can leads to slow-roll inflation.
The potential \eqref{potential1} has a stationary point at $\phi_c=\ln1+\frac{4\Lambda}{3m_0 ^2}$, this point correspond to 
minimum of potential $V''(\phi_c)>0$ if $m_0 ^2>0$. 
From potential \eqref{potential1}, we obtain the mass of a scalar state
\be
m_{\phi}^2=\frac{d^2V(\phi)}{d\phi^2}=\frac{9 m_0 ^4}{4 \Lambda +3 m_0 ^2}
\ee\label{mass2}
Through the above equation, like $m_{\phi} ^2=m_{0} ^2$, we find that the cosmological constant $\Lambda=\frac{3}{2}m_0 ^2$. 
Observationally we know that $\Lambda\sim (10^{-42} GeV)^2$ which is a very small value and we obtain that $m_0\sim 10^{-42}$ GeV which imply that the correction to Newton's law \eqref{new.pot} is very large.
\subsection{Toy model (case $n=1$)}
The objective of this section is to illustrate the results obtain in previous section for the following toy model corresponding to $n=1$ and $\epsilon=1$ in the original theory \eqref{BI}. This model has been previously studied in \cite{Kruglov}.
Through the eq. \eqref{trace1} we find 
\be
\frac{4 \alpha  R-4 \sqrt{\alpha  R+1}+4}{\alpha }-R=0
\ee
which is the algebraic equation with respect to $R$, the roots of this equation are
\be
R=0\quad R=-\frac{8}{9\alpha}
\ee
Like $\alpha>0$ the second solution corresponds to $AdS$ space.
\begin{figure}
\begin{center}
\includegraphics[height=2in,width=2in]{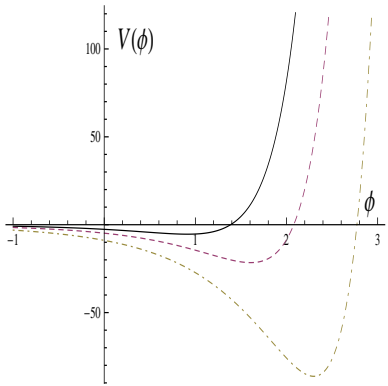}
\caption{Potential of the scalar field \eqref{potential2} for different choice of the parameter $\alpha$,  black line  $\alpha=1/2$, dashed line $\alpha=1/4$ and dotdashed line $\alpha=1/8$. }
\end{center}
\end{figure}\label{fig2}

We know that these kinds of modifications are plagued of several kind of instability.  
In this case through the eq. \eqref{Riccicond} we find that the theory is free Dolgov-Kawasaki Instability for
\be\label{RS1}
R_0\geq-\frac{1}{\alpha}
\ee
the effective gravitational coupling \eqref{EGC} is positive for 
\be\label{EGC1}
-\frac{1}{\alpha}<R<\frac{3}{2\alpha}
\ee

The conditions of Dolgov-Kawasaki stability \eqref{RS1} and positivity of the
effective gravitational coupling \eqref{EGC1} together imply absence of tachyon
instability for
\be
-\frac{1}{\alpha}<R<\frac{3}{2\alpha}
\ee
The condition $f''(R)\geq0$ is warranty that the scalar degree of freedom is not a ghost.
Through \eqref{1.2} for  $n=1$ and $\epsilon=1$ we find 
\be
e^{-\phi}=\frac{\alpha }{2 \sqrt{1+\alpha R}}
\ee
The self-interaction potential of the scalar field has the following form
\be\label{potential2}
V(\phi)=\frac{e^{\phi } \left(\alpha  e^{\phi }-2\right) \left(3 \alpha  e^{\phi }+2\right)}{4 \alpha }
\ee
In the Fig.2 we show the potential \eqref{potential2} and we can see how $V(\phi)\rightarrow0$ if $\phi\rightarrow-\infty$ while for $\phi\rightarrow\infty$ the potential diverge.
\be
V'(\phi)=\frac{e^{\phi } \left(\alpha  e^{\phi } \left(9 \alpha  e^{\phi }-8\right)-4\right)}{4 \alpha }
\ee

The potential \eqref{potential2} has a stationary point at $\phi_c=\ln\frac{2(2+\sqrt{13})}{9\alpha}$, in this point \eqref{potential2} possesses minimum $V''(\phi_c)>0$.
From potential \eqref{potential2}, we obtain the mass of a scalar state
\be\label{mass3}
m_{\phi}^2=\frac{d^2V(\phi)}{d\phi^2}=\frac{27}{4} \alpha  e^{3 \phi }-\frac{e^{\phi }}{\alpha }-4 e^{2 \phi }
\ee
We can consider that present universe corresponds to the minimum  $\phi_c$ of $V(\phi)$.
It follows from \eqref{mass3} that $m_{\phi}^2=\frac{4 \left(52+17 \sqrt{13}\right)}{81 \alpha ^2}$.  If the constant $\alpha$ is very small the effective mass $m_{\phi}^2$ is large and the scalar field  $\phi$ decouples, hence corrections to  Newton's law \eqref{new.pot} are negligible.  Then the obtained theory does not conflict with the cosmological observations, say, the solar system observations \cite{Will2001}.

In \cite{Salcedo2010a} was studied the phase space of a theory like this with $\epsilon=-1$ and $n=1$ and was proved that this model can produce late time acceleration, which is associated with a solution dominated by the curvature see \cite{Salcedo2010a}.

\section{Discussion}

In summary, in this paper we have considered a modification type Born-Infeld to Einstein-Hilbert action. We paid special attention to the issues associated with linearization of this theory around vacuum, maximally symmetric spaces of constant curvature.  

The gravitational spectrum of the linearized theory show the existence massless graviton plus a new scalar degree of freedom of spin-$0$ which is associated with $R^2$-term in \eqref{BIlinea}. The existence of this new scalar degree of freedom modified the newtonian potential, however for large values of $m_0$ this correction is negligible. 
We prove that the masses of the scalar degree of freedom  corresponding to original theory and linearized theory are equals. 

From the potential \eqref{potential1} corresponding to linearized theory we find that the mass of the scalar field is of order $m_0\sim 10^{-42}$ GeV which imply that the correction  to Newton's law \eqref{new.pot} is not negligible. The potential \eqref{potential1} behaves almost constant when $\phi\rightarrow\infty$ which can leads to inflation in the early universe.

The toy model is free of the Dolgov-Kawasaki instability, Tachyon instability  and negative effective gravitational
coupling for $-\frac{1}{\alpha}<R<\frac{3}{2\alpha}$. The condition $f''(R)\geq0$ is warranty that the scalar degree of freedom is not a ghost.   Also it was demonstrated that for small values of the constant $\alpha$ the mass of scalar degree of freedom is large and  the corrections to Newton's law are negligible. Also  in \cite{Salcedo2010a} was proved that this model can leads to a late time acceleration state in the universe without to include Dark Energy, in this model the acceleration of the expansion is produced by nonlinear effects  of the curvature.

\bibliography{bib}
\bibliographystyle{apsrev}

\end{document}